\documentclass[%
 reprint,
 amsmath,amssymb,
 aps,
]{revtex4-2}

\usepackage{graphicx}
\usepackage{dcolumn}
\usepackage{bm}
\usepackage{placeins}
\usepackage{comment}
\usepackage{color}

\newcommand{\corr}{\textcolor{black}}

\begin{document}


\title{Multipole analysis of substrate-supported dielectric nanoresonator metasurfaces with T-matrix method }

\author{Krzysztof M. Czajkowski}
\email{krzysztof.czajkowski@fuw.edu.pl}
\affiliation{Faculty of Physics, University of Warsaw, Pasteura 5, 02-093, Warsaw, Poland}
\author{Maria Bancerek}
\affiliation{Faculty of Physics, University of Warsaw, Pasteura 5, 02-093, Warsaw, Poland}
\author{Tomasz J. Antosiewicz}
\email{tomasz.antosiewicz@fuw.edu.pl}
\affiliation{Faculty of Physics, University of Warsaw, Pasteura 5, 02-093, Warsaw, Poland}

\date{\today}

\begin{abstract}
Substrates, and layered media in general, are ubiquitous and affect the properties of any object in their vicinity. However, their influence is, in an arbitrary framework, challenging to quantify analytically, especially for large arrays which additionally escape explicit numerical treatment due to the computational burden. In this work, we \corr{utilize} a versatile T-matrix based framework \corr{to} generalize the coupled multipole model towards arbitrarily high multipole orders and substrate-supported arrays. We then employ it to study substrate-supported random/amorphous arrays of high index dielectric nanoparticles which are of wide interest due to relatively low losses and a highly tunable optical response, making them promising elements for nanophotonic devices. We discuss how multipole coupling rules evolve in the presence of a substrate in amorphous arrays for three interaction mechanisms: direct coupling between particles, substrate-mediated interparticle coupling and substrate-mediated self-coupling. We show how the interplay of array density, distance from the substrate and its refractive determine the optical response of an array. As an example, we use this framework to analyze refractometric sensing with substrate-supported arrays and demonstrate that the substrate plays a crucial role in determining the array sensitivity.
\end{abstract}

\maketitle


\section{Introduction}
\noindent
High index dielectric (HID) nanoresonators are used as building blocks of novel photonic devices such as photonically enhanced photovoltaic cells, biomolecule sensors and flat analogues of conventional optical devices called metasurfaces \cite{Staude2017}. The interest in HID nanoresonators stems from the fact that they support both electric and magnetic resonances in simple geometries such as spheres or disks, what provides significant tunability of the optical response \cite{Kuzntesov2012}. Also, they are less susceptible to losses and they are more compatible with the CMOS standard of modern electronics than plasmonic counterparts \cite{Staude2017}.

A convenient way to express and analyze the fields scattered by HID nanoresonators is by multipole expansion. The far-field response of an antenna is related to the interference of its multipole fields, the manipulation of which can, for example, lead to unidirectional scattering via generalized Kerker effects \cite{Liu2018}. Tailored, directional scattering is essential for nonlinear photonics with HID nanoresonators and design of Huygens metasurfaces, which exploit Kerker effects to obtain almost ideal Huygens sources \cite{Decker2015}. The conditions for achieving directional scattering are almost exclusively expressed using multipole moments. Notably, while small particles are usually associated with dipole moments, higher-order multipoles are also important and, upon careful design of nanoresonator geometry, can be even dominant in its scattering spectrum \cite{Zenin2020}.

Antennas can, of course, be assembled into arrays. Their optical response is then determined by an interplay between the single-particle response, multiple scattering, and interference of the emitted fields. Multiple scattering leads to radiative coupling of multipole moments of nanoresonators and is addressed by solving a set of self-consistent equations. 
The coupled multipole approach retains the physical interpretation of the inner working of an array and begets an intuitive understanding of the properties, such as shifting and broadening of antenna resonances due to near-field coupling, or lattice resonances in periodic arrays \cite{PRL_101_143902_barnes}. It also demonstrates that different multipoles are affected in distinct manners by tuning the periodicity of an array in orthogonal directions. By virtue of this fact, electric and magnetic dipole resonance wavelengths can be tuned independently, modifying the resonance overlap condition \cite{Babicheva2018a}. 

\corr{Alternatively, interparticle coupling can be shaped by using amorphous arrays, which are random with a constraint on the minimal separation between nanoresonators. This constraint introduces short range position correlation and long range disorder. The randomness eliminates lattice resonances, leaving the optical response qualitatively similar to that of a single nanoresonator. In general, the minimal center-to-center (CC) distance between nanoresonators in an amorphous arrays changes interparticle coupling and the optical response. These modifications manifest themselves as, for example, changes of the resonance wavelength and quality factor \cite{Antosiewicz2012}, scattering-to-absorption ratio \cite{Antosiewicz2015}, or directional scattering and solar energy harvesting efficiency \cite{Czajkowski2020a}.}

The versatility of the coupled multipole model stems from that it provides semi-analytical solutions for the multipole moments of infinite nanoresonator arrays in both periodic and amorphous arrangements \cite{Czajkowski2020a}. To that end, first, propagators of each multipole must be derived and a general coupled multipole equation system must be proposed. Then, an assumption that the nanoparticle array is infinite is introduced. For periodic arrays this renders the multipole moments of all particles identical and reduces the inverse problem to single particle multipoles coupled to their infinite multipole neighbourhood via so-called lattice sums. The procedure is exactly the same for amorphous arrays \cite{Antosiewicz2012,Czajkowski2020}, except for the fact that in random arrays each nanoparticle has a unique neighbourhood and therefore an average, continuous multipole film is considered as the nanoparticle's neighbourhood. In recent literature, there are several examples of lattice sum derivations including electric and magnetic dipoles, dipole-quadrupole coupling for both periodic \cite{Babicheva2019, Terekhov2019} and amorphous arrays \cite{Antosiewicz2012, OE_22_2031_anto, Czajkowski2020, Czajkowski2020a}.

A substantial disadvantage of the coupled multipole model for nanoresonator arrays is its applicability to a homogeneous environment.
However, two-dimensional nanoresonator arrays are almost exclusively fabricated on a substrate. One prominent example of substrate-related effects is exceptional field enhancement observed in a nanoparticle-on-mirror system , in which a plasmonic or a HID nanoresonator is placed in close vicinity of a metallic mirror \cite{Sugimoto2018, Maimaiti2020}. From a mathematical point of view, the presence of a substrate influences the multipole expansion of the scattered fields \cite{Butakov2016, Chen2017}.
For example, the scattered electric dipole field can be reflected off the substrate and trigger a magnetic dipole response and vice-versa, leading to magnetoelectric coupling \cite{Miroshnichenko2015} and substrate-induced bianisotropy \cite{Albooyeh2015}. Magnetoelectric coupling also leads to exceptionally strong polarization sensitivity of the optical response of HID nanoresonators placed on a metallic film \cite{Sinev2016}. Finally, the presence of a substrate is known to modify the back-reflection Kerker conditions \cite{Pors2015, Babicheva2017} and circular dichroic spectrum of a nanoresonator \cite{Nechayev2019, Klos2019, Garcia-Guirado2020}.

In this work, we exploit recent advantages in the transition matrix (T-matrix) method to address the generalization of the coupled multipole model towards arbitrarily high multipole orders and substrate-supported arrays. The T-matrix is closely related to Cartesian multipole moments and the two formulations (superposition T-matrix and Green function based) of the coupled multipole model are equivalent \cite{Grahn2012, Muhlig2011}. Both require the evaluation of the so-called Sommerfeld integrals in order to calculate the reflected fields.
This approach has been used for instance to evaluate magnetoelectric coupling in a single dielectric particle \cite{Miroshnichenko2015} or substrate-supported single nanoresonator optical response in the discrete dipole approximation \cite{Schmehl1997}. Recently, a general superposition T-matrix method for nanoresonators in a layered medium has been proposed \cite{Egel2014}. 
\corr{As the multipole expansion of the field is used in this work, it is prudent to note that despite the usefulness of distinguishing between toroidal and non-toroidal multipole moments, such a separation cannot be done here. One should note that the effect of the toroidal moments is included in the scattering coefficients calculated based on the T-matrix. Lack of such distinction does not affect the physics of the interparticle coupling captured by the T-matrix formulation as the interparticle coupling cannot modify the toroidal and non-toroidal moments independently \cite{Evlyukhin2016}. This stems from the fact that the toroidal moments are actually higher-order corrections to well-known long-wavelength approximations to multipole moments \cite{Alaee2019}.}

Here, we formulate an infinite array approximation, which provides the effective (average) multipole moments of a nanoresonator in an array. We focus on amorphous arrays as they are challenging to tackle with other methods and are an interesting alternative to periodic ones, especially since many bottom-up, self-assembly methods exist to fabricate such random structures. However, the model and some of the conclusions are also applicable to periodic arrays. This work is structured as follows. First, we derive the model and verify it numerically with the superposition T-matrix and finite-difference time-domain (FDTD) methods. Then, we study multipole coupling in a substrate-supported array. We discuss general multipole coupling rules that can be applied to any array with central symmetry. We exemplify these rules and show generalized magnetoelectric coupling that includes higher-order multipoles. Finally, we study the parameters influencing substrate-mediated multipole coupling and disucss the physics of electromagnetic coupling effects in refractometric sensing.

\section{Amorphous array of particles on a planar substrate}
\noindent
The external $\bm{E_{ex}(r)}$ field acting onto a particle and its scattered $\bm{E_{sc}(r)}$ field at position $\bm{r}$ can be expanded into vector spherical wave functions (VSWFs) as follows
\begin{equation}
    \bm{E_{ex}}(\bm{r})=\sum_{l=1}^{\infty} \sum_{m=-l}^n a_{ml}^E \bm{M^1}_{ml}(k\bm{r})+a_{ml}^M \bm{N^1}_{ml}(k\bm{r}),
\end{equation}
\begin{equation}
    \bm{E_{sc}}(\bm{r})=\sum_{l=1}^{\infty} \sum_{m=-l}^n b_{ml}^E \bm{M^3}_{ml}(k\bm{r})+b_{lm}^M \bm{N^3}_{ml}(k\bm{r}),
\end{equation}
where $k=2\pi n_{m}/\lambda$ is the wavenumber in a given medium with index $n_{m}$, $\lambda$ is the wavelength, and we follow \cite{Doicu2014} in the definitions of VSWFs, which for convenience are summarized in Appendix \ref{ap:vswf}.
The T-matrix relates the expansion coefficients of the external ($\bm{a}$) and
scattered fields ($\bm{b}$) in terms of radiating VSWF \cite{Doicu2014}
\begin{equation}
\begin{pmatrix}
  b^{E}  \\ b^{M}
 \end{pmatrix} = \begin{pmatrix}
  T^{EE} & T^{EM} \\
  T^{ME} & T^{MM}
 \end{pmatrix}
 \begin{pmatrix}
  a^{E} \\ a^{M}
 \end{pmatrix},
\end{equation}
which is simplified as $\bm{b}=T\bm{a}$. Here, the T-matrices are calculated using the null-field method with discrete sources which is an efficient method of evaluating single particle scattering properties \corr{\cite{Doicu1999}}. 

In the case of coupled nanoantennas embedded in a stratified medium the equation describing the response of any scatterer $S$ is extended \cite{Egel2016} to include the scattered field of all other scatterers $S'$ as
\begin{equation}
\bm{b^{S}}=T^S\left(\bm{a^{S}}+\sum_{S'\neq S} \bm{a^{S,S'}_{d}} + \sum_{S'} \bm{a^{S,S'}_{r}}\right).
\label{eq:generalcoupling}
\end{equation}
The scattered field from the other scatterers, $S'$, is expressed in regular rather than radiating VSWFs to conveniently consider this field as a contribution to the incident field driving scatterer $S$. The relation that defines the direct (subscript $d$) coupling matrix is therefore
\begin{equation}
    \bm{a^{S,S'}_{d}}=W^{S,S'}_{d}\bm{b^{S'}},
\end{equation}
where $W^{S,S'}_{d}$ is the direct coupling matrix which is also present in a homogeneous environment.
A similar expression is defined for the scattered field from scatterer $S'$ reflected (subscript $r$) off the substrate
\begin{equation}
    \bm{a^{S,S'}_{r}}=W^{S,S'}_{r}\bm{b^{S'}},
\end{equation}
where $W^{S,S'}_{r}$ is the substrate-mediated coupling matrix. The definitions of the coupling matrix terms are presented in Appendix \ref{ap:cpl-matrix}.

By combining the definitions of the coupling matrices with Eq. \ref{eq:generalcoupling} we obtain
\begin{equation}
\bm{b^{S}}=T^S\bigg(\bm{a^{S}}+\Big(\sum_{S'\neq S}W^{S,S'}_{d}+\sum_{S'}W^{S,S'}_{r}\Big)\bm{b^{S'}}\bigg).
\end{equation}
We separate out the reflected self-coupling of scatterer $S$ with itself ($W^{S,S}_{r}$) from substrate-mediated interparticle coupling ($W^{S,S'}_{r}$) and under the summation we explicitly introduce the dependence of $W_{d}^{S,S'}$ and $W_{r}^{S,S'}$ on cylindrical coordinates and drop the superscripts
\begin{multline}
  \bm{b^{S}}=T^S\Big(\bm{a^{S}}+ \Big[W^{S,S}_{r}+\sum_{S' \neq S} \big(W_{d}(\rho_{S,S'},\phi_{S,S'})+\\
  +W_{r}(\rho_{S,S'},\phi_{S,S'})\big)\Big]\bm{b^{S'}}\Big),
  \label{eq:before_avg}
\end{multline}
where $(\rho_{S,S'},\phi_{S,S'})$ is a vector in cylindrical coordinates from scatterer $S$ to $S'$.

\begin{figure}
    \centering
    \includegraphics{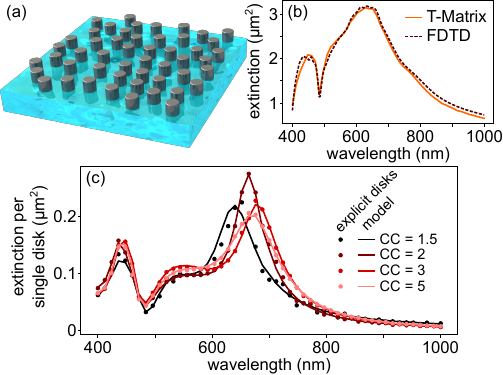}
    \caption{(Color online) (a) Small-scale schematic representation of substrate-supported random arrays of nanodisks (c-Si, $n_{sub}=2$, diameter $D=160$~nm, thickness $H=120$~nm) for comparison between FDTD and T-matrix calculations with (b) obtained extinction spectra showing very good agreement. (c) Comparison of T-Matrix calculations of explicit substrate-supported arrays of 1751 nanodisks with refractive index 4 [other parameters from (b)] with proposed model shows good agreement. Coupling in the random arrays with different density induces significant changes of the peak positions, widths, and amplitudes.}
    \label{fig:1}
\end{figure}

\corr{To reduce the many-particle problem to an effective single-particle one described with a self-consistent equation, we make use of the nature of amorphous arrays composed of identical scatterers, cf. Fig.~\ref{fig:1}a. Namely, while the neighborhood of any given resonator is unique, the fluctuations of the scattering coefficients of each neighbouring particle are sufficiently small to be neglected. Then, assuming an infinite array and averaging over different realizations of spatial disorder, it is possible to describe} the particle distribution in the array by a pair correlation function (PCF) $\Gamma(\rho/l_{cc},\phi)$\corr{, see Appendix \ref{ap:pcf}. $\Gamma$ is parametrized by the minimal center-to-center distance $l_{cc}=CC \times D$, where $D$ is the nanodisk diameter (with thickness $H$) and $CC$ is a dimensionless parameter.} This stochastic similarity of a random array allows us to replace the discrete particle properties $\bm{b^{S'}}$ by an effective, continuous \emph{film of multipoles} of average properties given by $\bm{b^{S}}$ and density by $\Gamma(\rho/l_{cc},\phi)$. Thus,
\begin{multline}
  \bm{b^{S}}=T^S\Big(\bm{a^{S}}+ \Big[W^{S,S}_{r}+  
  \sigma\!\int_{0}^{\infty}\!\!\rho d\rho\!\int_0^{2\pi}\!\!d\phi\Gamma(\rho/l_{cc},\phi) \times \\
  \times\big(W_d (\rho,\phi)+W_r(\rho,\phi)\big) \exp(-\varepsilon\rho)\Big]\bm{b^{S}}\Big),
  \label{eq:after_avg}
\end{multline}
where $\sigma$ is the particle number density \corr{and the passage from Eq. (\ref{eq:before_avg}) to Eq. (\ref{eq:after_avg}) is described in more detail in Appendix \ref{ap:avg}. For amorphous arrays, which are made using random sequential adsorption (RSA) \cite{Hinrichsen1986}, we have $\sigma=\sigma_0/l_{cc}^2$, where $\sigma_0\simeq0.696$ is a surface packing parameter which stems from the surface jamming limit \cite{2000_CollSurf_165_287_talbot}. As the radiative, proportional to $r^{-1}$ terms of the coupling matrices yield an improper integral without a well-defined limit for $r\to\infty$,  
we introduce an exponentially decaying term ($\exp(-\varepsilon\rho)$), where $\varepsilon$ a small constant equal to $10^{-5}$~nm$^{-1}$, to make the integrals well defined. Physically, this corresponds to an infinite array illuminated by a finite, slowly decaying beam.}

The equation can be further simplified by analyzing the angular integral. The coupling matrix can be factorized into radial and angular terms
\begin{equation}
    W_{r,d}(\rho,\phi)=W_{r,d}(\rho)\exp\big(i(m'-m)\phi\big).
\end{equation}
The angular probability distribution of finding a neighbouring particle is uniform, which leads to
\begin{multline}
  \bm{b^{S}}=T^S\Big(\bm{a^{S}}+ \Big[W^{S,S}_{r}+ 2\pi\sigma\int_{0}^{\infty}\rho d\rho\Gamma(\rho/l_{cc}) \times \\
  \times\big(W_{d}(\rho)+ W_{r}(\rho)\big)\exp(-\varepsilon\rho)\Big]\bm{b^{S}}\Big)
  \label{eq:effmedium}
\end{multline}
for $m_1=m_2$ and zero otherwise. Next, we define the effective direct coupling matrix as
\begin{equation}
 \widetilde{W}_d=2\pi\sigma\int_{0}^{\infty}\rho d\rho\Gamma\left(\rho/l_{cc}\right)W_{d}(\rho)\exp(-\varepsilon\rho),   
\end{equation}
and the effective substrate-mediated (reflected) coupling one as
\begin{equation}
 \widetilde{W}_r=2\pi\sigma\int_{0}^{\infty}\rho d\rho\Gamma\left(\rho/l_{cc}\right)W_{r}(\rho)\exp(-\varepsilon\rho).
\end{equation}

The radial integration is performed numerically, during which the improper integrals are truncated at a certain interparticle distance. There are three parameters that determine the accuracy of the integration: resolution of the radial grid, damping of the oscillating part of the integral, and truncation distance. Equation~\ref{eq:effmedium} is then solved for the scattering coefficients $\bm{b^S}$ by matrix inversion
\begin{multline}
    \bm{b^{S}}=\left[\big(T^S\big)^{-1}-\big(W^{S,S}_{r}+\widetilde{W}_d+\widetilde{W}_r\big)\right]^{-1}\bm{a^{S}}.
\end{multline}
Finally, the extinction cross-section is evaluated as a sum of individual VSWFs contributions \cite{Doicu2014}
\begin{equation}
    C_{ext}=-\frac{\pi}{k^2}\sum_{n=1}^{\infty} \sum_{m=-n}^n \mathrm{Re}\left(a_{nm}^{*E} b_{nm}^E+a_{nm}^{*M} b_{nm}^M\right).
    \label{eq:cext}
\end{equation}

Having outlined the model, we verify it numerically by calculating optical properties of  substrate-supported amorphous arrays of HID nanodisks illuminated by a normally incident plane wave. Modelling of anisotropic scatterers with the T-matrix method is problematic due to the Rayleigh hypothesis that states that the fields calculated with the T-matrix method are valid only outside the circumscribing sphere of the scatterer. In our particular case of HID nanodisks supported by a substrate, such a sphere crosses the substrate and therefore the Rayleigh hypothesis is not fulfilled for a particle interacting with its image. However, it has been recently shown that this limitation can be circumvented by relying on the conditional convergence of plane wave expansion of VSWFs truncated at a certain wave vector \cite{Egel2017}. Here, we utilize the relation between the truncation wave vector and the truncation multipole order from \cite{Egel2017} to simulate anisotropic nanoresonators. Because of the fact that full-wave Maxwell equation solvers based on finite-element or FDTD methods are not capable of simulating random arrays of realistic size, we validate our approach in two steps.

First, we confirm the validity of the plane-wave expansion for substrate-supported arrays by comparing the extinction cross-section spectrum obtained with the T-matrix method (open-source python code SMUTHI \cite{Egel2014, Egel2016}) for an amorphous array composed of 24 crystalline silicon (c-Si) \cite{Schinke2015} nanodisks ($D=160$ nm, $H=120$ nm, substrate index $n_{sub}=2$) with an FDTD simulation (FDTD Solutions, Lumerical) for the same array, cf. Fig.~\ref{fig:1}b.
Next, we compare the results of superposition T-matrix to our effective model. We simulate substrate-supported arrays with 1751 particles made of c-Si and assembled using RSA. This number of disks is sufficient for obtaining converged spectra for different realizations of an amorphous array. For low density (high $l_{cc}>10D$) the optical response is almost density independent \cite{OE_22_2031_anto}, as it should, since the arrays tend to well-separated quasi-single disks. However, as plotted in Fig.~\ref{fig:1}c, for dense arrays significant deviation from the single particle response is observed \cite{Antosiewicz2012, Czajkowski2020a}. Agreement of the two methods is maintained even for dense arrays, confirming that the effective approach calculates correct optical spectra of substrate-supported amorphous arrays orders of magnitude faster than the direct superposition T-matrix one.

\section{Multipole coupling selection rules}
\noindent
Occurrence of multipole coupling depends on three factors: on the coupling matrix form of a given particle environment, on symmetries of the angular distribution of particles and on the polarization of the external field (non-zero coefficients in the VSWF expansion $\bm{a}$) \cite{Doicu2014}. \corr{We now recall the multipole coupling rules for the free-space case emerging from the translation theorem for VSWF \cite{Pendry1974}. We then analyze the relaxation of those rules for substrate-supported scatterers and discuss its consequences for the interparticle coupling. }

Let us then consider the associated Legendre polynomial part of the direct coupling matrix $W_d^{S,S'}$,
\begin{equation}
    P_\chi^{|m-m'|}(\cos{\theta}),
\end{equation}
where 
$\chi \in \left[  |l-l'|, l+l'  \right] $ 
and $\theta$ is the azimuthal angle in spherical coordinates. Since interparticle coupling vanishes unless $m=m'$, one can use the recurrence relation
\begin{equation}
    P_{n+1}(\cos{\theta})=\alpha \cos{\theta} P_{n}(\cos{\theta}) - \beta P_{n-1}(\cos{\theta})
\end{equation}
with $P_0=\frac{\sqrt{2}}{2}$ and $P_1=\sqrt{\frac{3}{2}} \cos{\theta}$ to evaluate these polynomials. As a consequence of the fact that we are analyzing a planar array, we set $\cos{\theta}$ to zero. Then only even $\chi$ contribute to the overall result. 

The other factor governing the occurrence of coupling between multipoles of given degrees are Wigner-3j symbols, where 
\begin{equation}
    w^a(l,l',\chi)= 
\begin{pmatrix}
l & l' & \chi \\
0 & 0 & 0 
\end{pmatrix}
\end{equation}
contributes to coupling between multipoles of the same type (i.e. electric-electric or magnetic-magnetic) and
\begin{equation}
    w^b(l,l',\chi)= 
\begin{pmatrix}
l & l' & \chi-1 \\
0 & 0 & 0 
\end{pmatrix}
\end{equation}
contributes to cross-coupling between electric and magnetic multipoles.

Such Wigner-3j symbols vanish unless the sum of its top row is an even integer. Because the condition that stems from the Legendre polynomials limits $\chi$ to even numbers, we have that if $l+l'$ is even, then $w_a \neq 0$ and $w_b = 0$. Otherwise, if $l+l'$ is odd, then $w_a=0$ and $w_b \neq 0$. 
These conditions lead to the following conclusion: coupling between electric and magnetic multipoles can happen only if one of their orders is even and the other is odd. For such orders, coupling between multipoles of the same types does not occur. Otherwise, only coupling between the same multipole types occurs. This is a generalization of previously known examples of multipole coupling selection rules. 
The presented selection rules are valid for particles embedded in a homogeneous environment. When particles are deposited on a substrate, these rules can be violated as a consequence of substrate mediated coupling, allowing multipoles with the same order $m$ to couple regardless of their type and degree.

\begin{figure}
    \centering
    \includegraphics{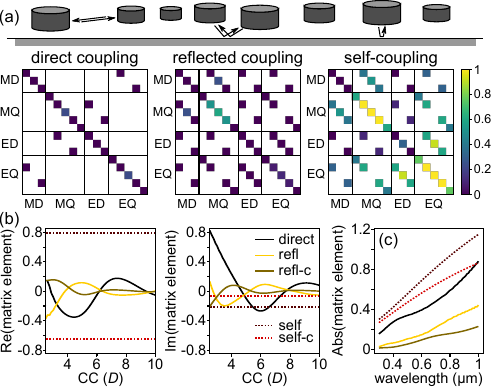}
    \caption{(Color online) (a) Coupling matrices for (left) direct coupling between particles, (center) substrate mediated interparticle coupling, and (right) substrate mediated self-coupling for CC $=3.5D$, $\lambda=700$~nm assuming $D=160$ nm and  $H=160$ nm. (b) Real and imaginary parts of the dipole matrix elements at $\lambda=700$~nm as a function of center-to-center distance for direct coupling (direct), substrate-mediated coupling between the same dipole type (refl) and cross-coupling between electric and magnetic dipoles (refl-c), and self-coupling also between the same dipoles (self) and cross-coupling (self-c). (c) The magnitude of the dipole matrix element as function of wavelength for CC $=3.5D$ shows that the self-coupling is the dominant factor modifying the optical properties, however, jointly the remaining contributions can be, depending on the relative phases, of similar magnitude.}
    \label{fig:2}
\end{figure}

The coupling matrix can be used as a tool for studying properties of multipole coupling between the particles as well as between the particles and the substrate (Fig.~\ref{fig:2}a). We distinguish three coupling types which originate from Eq.~\ref{eq:effmedium}: direct coupling ($\widetilde{W}_d$), substrate-mediated self-coupling ($W^{S,S}_{r}$) and substrate-mediated interparticle coupling ($\widetilde{W}_{r}$). Each of the coupling matrix terms can be considered individually to study the contribution of each mechanism into the final result, as presented in Fig.~\ref{fig:2}. The direct term, which is present even when the particle is embedded in a homogeneous environment, obeys the symmetry dictated by multipole coupling selection and is an extension of the dipole-dipole coupling \cite{OE_22_2031_anto, Czajkowski2020a}. An analysis of this analogy is presented in Appendix \ref{ap:relationship}. In contrast to direct coupling, both substrate-mediated coupling terms obey a less-strict selection rule providing a route towards observation of various multipole cross-coupling effects. The only rule except for $\Delta m = 0$ they obey, is that coupling is not possible between electric and magnetic multipoles when $m=0$. This rule is discussed in Appendix \ref{ap:cpl-matrix}.

\begin{figure*}
    \centering
    \includegraphics{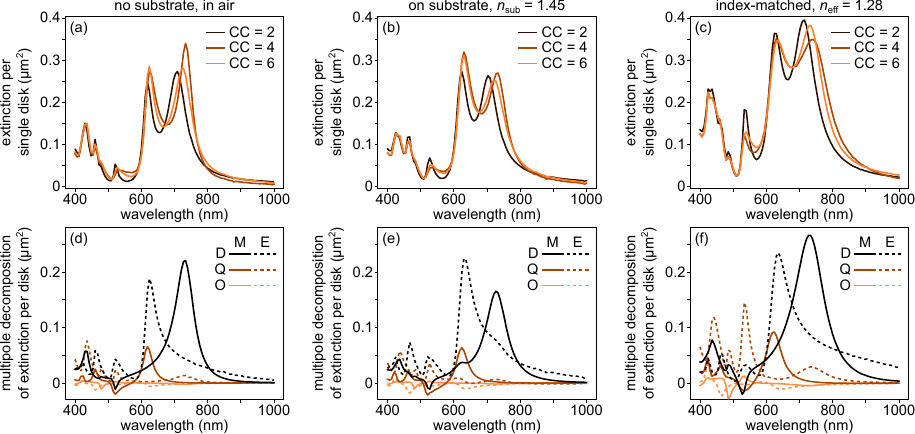}
    \caption{(Color online) Comparison between the optical response of amorphous arrays of c-Si cylinders with $D=150$~nm and $H=225$~nm for selected center-to-center distances: (a) in air, (b) on a substrate with $n_\mathrm{sub}=1.45$, and (c) in an index-matched medium of 1.28. (d-f) Multipole decomposition (D -- dipole, Q -- quadrupole, O -- octupole) of exctinction spectra for the three cases for CC = 5. Note the substrate mediated magneto-electric coupling around 600 nm and 725 nm, which is absent in homogeneous environments.}
    \label{fig:fig3}
\end{figure*}

To elucidate the influence of various parameters on multiple scattering leading to intra-array coupling, we now focus on dipolar terms only, which dominate the optical spectrum of nanodisks in the visible and near-infrared. While electric-electric and magnetic-magnetic dipole coupling is described by the same term, electric-magnetic dipole cross-coupling has its own unique dependence. As shown previously, (direct) interparticle coupling in amorphous metasurfaces depends on their density in an oscillatory manner of a stochastic, quasi-Fabry-Perot cavity \cite{Czajkowski2020a, Czajkowski2020}. Here, various coupling types also exhibit this distinct dependence, cf. Fig.~\ref{fig:2}b. While self-coupling does not depend on array density, substrate-mediated interparticle coupling, similar to direct coupling, does oscillate as a function of density, but its phase is shifted with respect to direct coupling. Similar oscillations occur also for interparticle cross-coupling between electric and magnetic dipoles, however, it is comparatively weaker than other coupling types, cf. Fig.~\ref{fig:2}c. 
In contrast, self-cross-coupling has nearly the same magnitude as self-coupling, but opposite sign (Fig.~\ref{fig:2}b). The magnitudes of the various dipole matrix elements as function of wavelength for $l_{cc}=3.5D$ show constant qualitative behavior, as plotted in Fig.~\ref{fig:2}c. In principle, the direct term is the one which has the biggest impact on the optical properties of a single substrate-supported particle. However, the remaining contributions can be of similar magnitude, especially for arrays of intermediate density with $CC$ around 4--7. 

We now study the optical exctinction spectra of substrate-supported amorphous arrays of HID nanodisks in the light of the above conclusions. First,  we  compare  the  response of metasurfaces composed of c-Si \cite{Schinke2015} cylinders with diameter of 150 nm and height of 225 nm in three environments  (see Fig. \ref{fig:fig3}a-c):  vacuum, on a substrate with $n_{sub}=1.45$, and an effective homogeneous medium with index 1.28. Qualitatively, dependence of the extinction spectrum on the minimal CC distance is the same for the vacuum and substrate-supported case. The difference between the two cases is most prominent for CC=4, for which the ratio between the magnetic and electric resonances is substantially different. In contrast, when the disks are placed in an effective medium, interparticle coupling is strongly modified as a consequence of the ratio between vacuum and medium wavelengths. Since reflection of waves scattered by the substrate is neglected, the phase relations of the scattered and external fields change significantly. Consequently, approximation of a substrate-supported case by using an effective permittivity is only poorly applicable to amorphous arrays.

Further differences between the optical properties of arrays embedded in a homogeneous and a layered medium can be observed by performing a multipole decomposition of the extinction spectra shown in Fig. \ref{fig:fig3}d-f. \corr{The contribution from each of the multipoles is obtained by summing only over the index $m$ in Eq.~\ref{eq:cext} with the type (electric and magnetic) defined by the superscript E/M}. The herein discussed disk supports spectral overlap of multiple electric and magnetic multipoles thanks to its dimension ratio $H/D>1$. This makes it an excellent candidate for observing magnetoelectric coupling and substrate induced bianisotropy.

\begin{figure}
    \centering
    \includegraphics{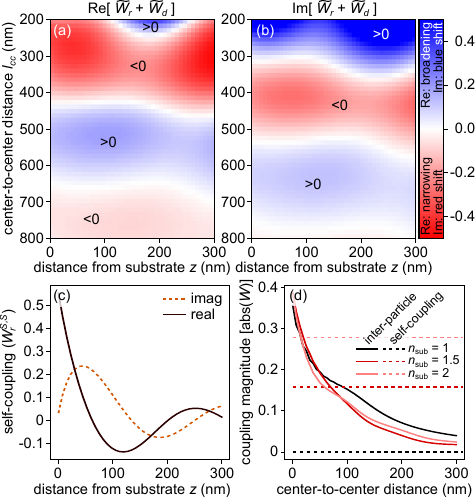}
    \caption{(Color online) Coupling dependence vs. distance from substrate and substrate index. (a) Real and (b) imaginary parts of the dipole-dipole coupling term $W$ (neglecting magneto-electric cross-coupling) for an amorphous array on a substrate ($n_\mathrm{sub}=2$) at $\lambda=500$ nm as function of array-substrate distance $z$ and minimum center-to-center $l_{cc}$ distance. The distance $z=0$ corresponds to an array of disks of height equal to 100 nm placed on a substrate. Both real and imaginary parts of total interparticle coupling shows distinct dependence on $l_{cc}$ for each array-substrate distance. (c) The self-coupling term $W^{S,S}_{r}$. likewise affected. (d) $W^{S,S}_{r}$ increases significantly with $n_\mathrm{sub}$, while interparticle coupling is weakly dependent on $n_\mathrm{sub}$ and only for very dense arrays with small $l_{cc}$ is larger than $W^{S,S}_{r}$.}
    \label{fig:fig4}
\end{figure}

First of all, the decomposition shows that \corr{for the substrate-supported case the electric dipole extinction spectrum has a characteristic shoulder at the magnetic dipole (Fig.~\ref{fig:fig3}e). It is indicative of mutual coupling  \cite{Miroshnichenko2015} and is markedly absent in both homogeneous environments (Fig.~\ref{fig:fig3}d,f)}. Furthermore, our calculations show that the presence of the substrate suppresses MD/EQ coupling at the magnetic resonance, which is inherently present in a homogeneous environment. 
The second significant resonance centered around 600 nm is composed of an electric dipole and a magnetic quadrupole, which is one of the consequences of $H/D>1$. 
Indeed, in contrast to the homogeneous cases, we observe an enhancement of the magnetic dipole around the composite ED/MQ resonance, in contrast to other works which reported even negative MD extinction in the vicinity of the ED resonance \cite{Miroshnichenko2015, Sinev2016}, as well as significant amplification of the ED resonance itself. In fact, all first four multipoles are coupled, as evidenced by a weak EQ peak around 600 nm. We attribute this observation to substrate-mediated multiple multipole coupling effects, which are observed, despite that fact that we use a relatively low-index substrate. Finally, we note that for the index matched case, in addition to the incorrect broadening of the dipolar resonances, this approximation erroneously predicts a sharp EQ around 500~nm.

\section{Factors influencing multipole coupling}

\begin{figure}
    \centering
    \includegraphics{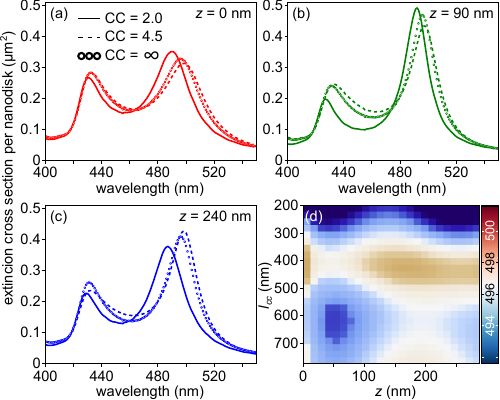}
    \caption{(Color online) Extinction cross-section spectra for silicon nanodisks with 100 nm diameter and 100 nm height placed on a substrate with $n_\mathrm{sub}=2$ for selected center-to-center distances and (a) $z=0$ nm, (b) $z=90$ nm, and (c) $z=240$ nm. The modification of the resonances is determined by the array density, which is inversely proportional to CC, and the array--substrate separation distance. \corr{(d) Example of magnetic resonance dependence on the disk's distance from the substrate $z$ and $l_{cc}$, which results from an interplay of various coupling terms. White marks the resonance wavelength of a single disk in free space.}}
    \label{fig:fig5}
\end{figure}

For an amorphous array of nanoparticles in free space, interparticle coupling depends on the minimal CC distance \cite{Czajkowski2020a}, while the presence of the substrate begets additional coupling mechanisms, which are governed by two factors. One is the substrate refractive index, which modifies the reflection coefficient and determines the strength of substrate-mediated coupling and the phase relationship between the field incident onto the substrate and the one reflected off it. The second one is the the array--substrate distance, which modifies the phase factors of substrate mediated coupling and its decay. Consequently, a substrate changes both interparticle- and self-coupling. To analyze this effect, we calculate total interparticle coupling, as a sum of  direct and reflected coupling terms, and the self-coupling between dipoles, in both cases neglecting higher-order terms and magnetoelectric coupling, as function of the array--substrate and minimal CC distances. In Fig.~\ref{fig:fig4}a-c we present the analysis for $\lambda=500$~nm, \corr{approximately} the magnetic dipole resonance wavelength for a silicon nanodisk with $H=100$~nm and $D=100$~nm in free space.

As shown in Fig. \ref{fig:fig4}a-b, the array--substrate distance modifies the dependence of the total interparticle coupling term on the minimal CC distance, especially for dense arrays. For the highest density analyzed ($l_{cc}$=200 nm), both the amplitude and phase of the coupling term can be tuned within the broad range. At the same time, for dilute arrays, the overall interparticle coupling strength is low and thus cannot be easily modified by changing the phase term via placement of an array. However, even then the optical response of an array depends strongly on its distance from the substrate due to the self-coupling term, as plotted in Fig. \ref{fig:fig4}c.  A similar modification of the self-coupling term can be observed when the refractive index of the substrate is changed, despite the fact that interparticle coupling is not significantly distinct from the free-space case, as shown in Fig. \ref{fig:fig4}d. Notably, the optical response of an array is always determined by the sum of self-coupling and interparticle coupling terms, which makes the array-substrate distance an important parameter for shaping how the interparticle coupling contributes the overall response of an array.  \corr{In general, the real and imaginary parts of $W$ affect, respectively, the width and position of the resonance. When Im$[W]>0$ and affected peak will blue shift, while for Re$[W]<0$ it will red shift. Simultaneously, a positive real part of $W$ broadens the resonance (and decreases its amplitude), while a negative one makes it more narrow (increasing the amplitude).}

\begin{figure*}[t]
    \centering
    \includegraphics{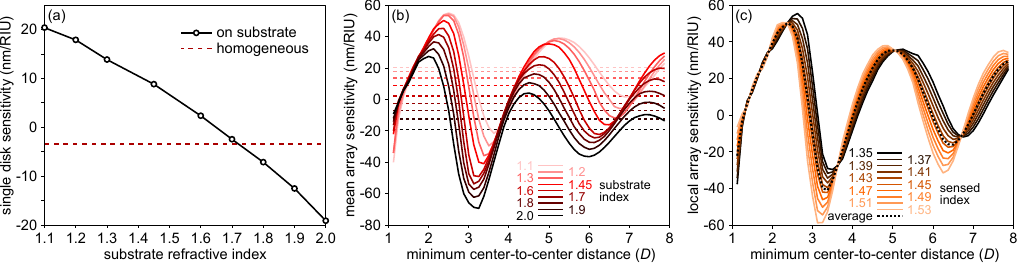}
    \caption{(Color online) Bulk sensing ($n_{d}\in\langle1.33,1.53\rangle$) characteristics of substrate supported dielectric nanodisks. (a) Sensitivity of a nanodisk in a homogeneous environment (dashed line) is very small, but when the disk is placed on a substrate (solid with circles) with $n_\mathrm{sub}\in\langle1.1,2\rangle$ it varies significantly with $n_\mathrm{sub}$. (b) The average sensitivity (over the whole $n_d$ range) of substrate-supported random arrays of disks (solid lines) varies with the center-to-center distance and the mean single-particle value (dashed lines) around which it oscillates depends on $n_\mathrm{sub}$. (c) The local-index ($\partial\lambda/\partial n_{d}$) sensitivity for substrate $n_\mathrm{sub}=1.45$ is non-linear and there is considerable spread of its value once near-field coupling effects have decayed for the center-to-center distance above $\sim2.5D$. }
    \label{fig:sensing}
\end{figure*}

To exemplify this we study the optical spectra of arrays composed of c-Si disks with $D$ and $H$ of 100~nm placed above a substrate with $n_{sub}=2$. The extinction spectra for selected array--substrate distances $z$ and minimal CC distances $l_{cc}$ are presented in Fig. \ref{fig:fig5}a-c. Both the amplitude and resonance wavelength are modified by changing the array--substrate distance, being consistent with the coupling matrices of Fig.~\ref{fig:fig4} for the magnetic dipole at 500~nm. For example, for a dense array with $CC=2$, the magnetic resonance increases in amplitude when going from $z=0$~nm to $z=90$~nm and then diminishes for $z=240$~nm. This is accompanied by corresponding, but inversely proportional width changes and concomitant peak shifts, whose evolution is offset in phase with respect to that of the peak amplitude \cite{Antosiewicz2012, OE_22_2031_anto}. The particular behavior of the peak amplitude stems from the fact that the real part of the interparticle coupling term has an opposite sign for $z=240$~nm in comparison with the two other cases.  The same conditions cannot be straightforwardly applied to the electric dipole resonance because of the different phase relationship within the coupling matrix between the resonance wavelength and the array--substrate distance. For $CC=4.5$ the properties are close to that of a single particle, yet each array-substrate distance results in a distinct modification of the optical response due to the relation between the coupling matrix and array-substrate distance. \corr{Figure~\ref{fig:fig5}d summarizes these observations for the magnetic dipole resonance position, whose dependence on $z$ and $l_{cc}$ results from direct $\tilde{W}_{d}$ and reflected $\tilde{W}_{r}$ interparticle coupling (Fig.~\ref{fig:fig4}b), self-coupling $W^{S,S}_r$ (Fig.~\ref{fig:fig4}c), and other terms like cross-coupling.}

\section{Refractive index sensing with HID antennas}

Amorphous arrays of HID nanodisks have recently been investigated in the context of refractometric biosensing  \cite{2017_Nanoscale_9_4972_bontempi, 2017_NL_17_4421_yavas}. Potential benefits of Si structures are low losses and high CMOS compatibility, providing a valid alternative for plasmonic sensors despite a having a lower sensitivity than plasmons \cite{Bosio2019}. Hence, unraveling the role various effects play in determining their response is important \cite{Yavas2019}. 
One factor enabling good sensitivity of an array of dielectric nanodisks is interparticle coupling due to an isolated one's low sensitivity in a homogenous environment \cite{Czajkowski2020a}. Thus, it is important to understand and quantify how a substrate affects its sensing characteristics: (i) by modification of the Fresnel coefficient as a function of the environment's refractive index and (ii) by modifying the phase factor of reflected waves.

We exemplify this by calculating bulk refractive index sensitivity for $n_d\in\langle1.35;1.55\rangle$ for a c-Si nanodisk with $D=H=160$~nm supported by a substrate with a varying $n_{sub}$ and compare it in Fig.~\ref{fig:sensing}a to a homogeneous environment. The sensitivity is calculated as the slope of a linear fit to the peak shift vs $n_d$. Indeed, the value of $n_{sub}$ substantially affects the sensitivity showing that even for an isolated particle it is a key factor. In principle, for a large contrast between the substrate and the medium the sensitivity is large, while for low contrast the sensitivity can even tend to zero. Moreover, the sign of peak shift depends on whether $n_{sub}$ is higher or lower than $n_d$ of the sensed medium.

When placed in an array, the optical response of the system is additionally modified by the interparticle coupling, which is an important factor determining the refractive index sensitivity of arrays embedded in a homogeneous environment \cite{Czajkowski2020a}. Next, we calculate the sensitivities of amorphous arrays of Si nanodisks with $D=160$ nm and $H=160$ nm for selected $n_{sub}$ as function of the minimum CC distance, which are plotted in Fig.~\ref{fig:sensing}b. The oscillating behavior of the sensitivity mirrors that of the interparticle coupling, which follows a damped, periodic function of $l_{cc}$.  The choice of the minimal CC distance is essential for obtaining high refractive index sensitivity \corr{through maximizing the imaginary part of the interparticle coupling matrix $W$} and maximizing the single particle response by coupling to an appropriate substrate, cf. Fig. \ref{fig:sensing}a. Consequently, the choice of substrate determines the sensitivity in the low density limit and then sets the potential range of values attainable by tuning the array density (minimum CC distance). For small $n_{sub}$ the optimal $l_{cc}$ is at the first maximum that occurs at $CC\approx2.5$ and gives approximately 55~nm/RIU (refractive index unit) sensitivity. In contrast, for $n_{sub}=2$ the best sensitivity occurs at $CC\approx3$ and while negative, it gives the global maximum (in the tested range) of 70~nm/RIU.

A further consequence of the oscillatory behavior of the minimal CC dependence of interparticle coupling is a nonlinear relation between $n_d$ and the wavelength shift.  If $n_d$ changes are small, the sensitivity is linear, however, not for large index changes. We illustrate this via the local sensitivity as $\partial\lambda/\partial n_d|_{n_d}$ and study this property for selected $n_d$ as a function of $l_{cc}$ for a fixed $n_{sub}=1.45$. The results are plotted in Fig. \ref{fig:sensing}c. In general, the observed local sensitivities are close to the average value (plotted also in Fig. \ref{fig:sensing}b). However, for $CC\ge2.5$, for which near-field effects have decayed, a substantial spread of the sensitivity values is observed, indicating the nonlinear dependence of the wavelength shift vs the refractive index of the environment. The effect is especially pronounced close to sensitivity minima.

\section{Conclusions}
In this work we introduced a computationally efficient and accurate T-matrix-based effective model for describing the optical properties of substrate-supported random arrays of nanoresonators, accounting for multiple scattering effects and the presence of the substrate. The advantage of the T-matrix framework is its close relationship with multipole decomposition, which we exploit to discuss general multipole coupling rules for both free-space and substrate-supported arrays. The model can be used to study the optical properties up to arbitrarily high multipole orders, which is advantageous with respect to a Green function based approach, where including each new multipole order requires an involved derivation of multipole propagators. 
Our approach extends beyond current multipole studies of substrate-supported arrays, which consider the dipole approximation in a decoupled case with substrate-mediated coupling neglected \cite{Babicheva2017}. We show how an important factor substrate-mediated coupling is and introduce generalized substrate-induced magnetoelectric coupling beyond magnetic dipole-electric dipole coupling.


We anticipate that this work will be useful for studying optical properties of large-scale photonic systems which contain nanoparticles and nearby interfaces. The proposed T-matrix based effective model can be used as a general and very efficient numerical tool for simulating substrate-supported nanoresonator arrays composed of identical particles with at least an approximate point symmetric spatial distribution. This includes not only amorphous but also periodic arrays, which are often present in metasurfaces. The coupling matrix and multipole decomposition are easily obtained from the model and can shine light onto contributions from interparticle and substrate-mediated coupling effects, which provide further insight into the properties of the coupled nanoparticle array-layered medium system.  

\appendix
\section{Vector spherical wave functions \label{ap:vswf}}
\noindent
In the T-matrix method the fields are expanded into regular and radiating vector spherical wave functions defined in the spherical coordinate system $(r,\theta,\phi)$ as
\begin{multline}
  \bm{M_{ml}}^{1,3}(k\bm{r}) = \sqrt{D_{ml}}z_l^{1,3}(kr)\left[im\frac{P_l^{|m|}(\cos{\theta})}{\sin{\theta}}\bm{e_{\theta}}- \right.\\
  \left.\frac{d}{d\theta}P_l^{|m|}(\cos{\theta})\bm{e_{\phi}}\right]e^{im\phi},
\end{multline}
\begin{multline}
  \bm{N_{ml}}^{1,3}(k\bm{r}) = \sqrt{D_{ml}}\left(\frac{l(l+1) z_l^{1,3}(kr)}{kr}P_l^{|m|}(\cos{\theta})\bm{e_{r}}+\right.\\ 
  \frac{\frac{d}{d(kr)}kr z_l^{1,3}(kr)}{kr}\left[\frac{d}{d\theta}P_l^{|m|}(\cos{\theta})\bm{e_{\theta}}+\right.\\
  \left.\left.im\frac{P_l^{|m|}(\cos{\theta})}{\sin{\theta}}\bm{e_{\phi}}\right]\right)e^{im\phi},
\end{multline}
where index 1 corresponds to regular VSWFs and 3 corresponds to radiating VSWFs, $(\bm{e_r},\bm{e_\theta},\bm{e_\phi})$ are the unit vectors in spherical coordinates, $z_l^{1,3}$ corresponds to the spherical Bessel $j_l$ and spherical Hankel functions of the first kind $h_l$ which correspond to superscripts 1 and 3, respectively, $P_l^{|m|}$ is the associated Legendre polynomial of order $l$ and $m$ and $D_{ml}$ is a normalization constant equal to
\begin{equation}
    D_{ml}=\frac{(2l+1)(l-|m|)!}{4l(l+1)(l+|m|)!}.
\end{equation}

\section{Layer-mediated and direct coupling matrices \label{ap:cpl-matrix}}
\subsection{Layer-mediated coupling}
\noindent
The layer-mediated coupling matrix is defined as \cite{Egel2016}
\begin{multline}
    W_{r,n,n'}^{S,S'}=4i^{|m-m'|}e^{i(m-m')\phi_{S,S'}} \times \\
    (I_{n,n'}^{+}(\rho_{S,S'},z_S+z_{S'})+I_{n,n'}^{-}(\rho_{S,S'},z_S-z_{S'})),
\end{multline}
where $n$ is a single index denoting the corresponding VSWF, which otherwise would require three indices $(l,m,\eta)$ with $\eta$ denoting magnetic ($0$ or $M$) and electric ($1$ or $E$) multipoles. $I_{n,n'}^{+}(\rho_{S,S'},z_S+z_{S'})$,$I_{n,n'}^{-}(\rho_{S,S'},z_S+z_{S'})$ are the Sommerfeld integrals meaning that the integral over the angular extent of the array is equal $2\pi$ for multipoles of with the same $m$ value and zero otherwise. 

To show how this matrix is constructed we refer to a formula from \cite{Egel2016} in which the Sommerfeld integral theorem is yet to be applied,
\begin{multline}
    W_{r,n,n'}^{S,S'}=\frac{2}{\pi}\sum_j \int \frac{d^2\bm{k_{||}}}{k_z k} e^{i (m'-m)\phi_{S,S'}}e^{i \bm{k_{||}}\cdot (\bm{r_S}-\bm{r_{S'}})}\times \\
    \times
    \begin{pmatrix}
B_{n,j}^{\dagger}(k_z/k) e^{i k_z z_S} & B_{n,j}^{\dagger}(-k_z/k) e^{-i k_z z_S}
\end{pmatrix} L(k_z) \\
\begin{pmatrix}
B_{n',j}(k_z/k) \\
B_{n',j}(-k_z/k)
\end{pmatrix},
\end{multline}
where
\begin{multline}
    B_{n,j}(x)=\frac{1}{i^{l+1}} \frac{1}{\sqrt{2l(l+1)}}(i \delta_{j,1}+\delta_{j,2}) \sqrt{1-x^2} \times \\
    \times \left(\delta_{\eta j}\frac{\partial P_l^{|m|}(x)}{\partial x}+(1-\delta_{\eta j})m\frac{P_l^{|m|}(x)}{1-x^2}\right),
\end{multline}
where $k_\parallel$ is the in-plane wave vector and $k_z$ is the wave vector component perpendicular to the substrate. $L(k_z)$ denotes the layer response matrix constructed according to \cite{Egel2014} and $j$ corresponds to summation over polarizations. $B^{\dagger}$ is defined as $B$ with all explicit $i$ substitued by $-i$.
For a simple case of a particle above a plane the layer-mediated coupling matrix reads
\begin{multline}
    W_{r,n,n'}^{S,S'}=\frac{2}{\pi}\sum_j \int \frac{d^2\bm{k_{||}}}{k_z k} e^{i (m'-m)\phi_{S,S'}}e^{i \bm{k_{||}}\cdot (\bm{r_S}-\bm{r_{S'}})} \times \\
    \times r(k_z) B_{n,j}^{\dagger}(k_z/k) B_{n',j}(-k_z/k) e^{ikz_S},
\end{multline}
as the substrate converts downward plane waves into upward reflected ones, the amplitude of which is modified by the Fresnel reflection coefficient $r(k_z)$. 

For $m=0$ the equation
\begin{equation}
    B_{n,j}(x) \propto \delta_{\eta j}\sqrt{1-x^2}\frac{\partial P_l^{|m|}(x)}{\partial x}.
\end{equation}
Consequently, if $\eta=\eta'$ and $m=m'=0$, then one of the terms of the sum in the integral defining $W$ is zero, however, the other one is not, as the product $B_{n,j}^{ \dagger}(k_z/k) B_{n',j}(-k_z/k)$ consists of two terms corresponding to the same plane wave polarizations. In contrast, if $\eta$ does not equal $\eta'$, then $B_{n,j}^{\dagger}(k_z/k) B_{n',j}(-k_z/k)$ consists of terms corresponding to the opposite polarizations, one of which always leads to zero and therefore the magnetoelectric coupling can never occur if $m=m'=0$. The same result can be provided  for the direct coupling by analyzing the Wigner-3j symbols in a similar manner to the one provided in the main text.

\subsection{Direct coupling}
\noindent
The direct part contains three factors: Bessel functions dependent on the product $kr_{S,S'}$, Legendre polynomials and $a_5$ and $b_5$ coefficients, which depend on Wigner-3j symbols. Here, we write those coefficients explicitly, because we study the properties of the Wigner-3j symbols in the main text. Due to the presence of the Legendre polynomial, it is convenient to define $W_d^{S,S'}$ in spherical coordinates $(r,\phi,\theta)$. If $\eta=\eta'$
\begin{widetext}
\begin{multline}
   W_{d,n,n'}^{S,S'}=e^{i(m-m')\phi_{S,S'}}{(-1)}^{m+m' }\frac{1}{4} \sqrt{\epsilon_{m}\epsilon_{m'}} 
   {\sum}_{\chi =|l-l' |}^{l+l' }{(-1)}^{(l' -l+\chi )/2}(2\chi +1) 
   \sqrt{\frac{(2l+1)(2l' +1)(\chi -(m-m' ))!}{l(l+1)l' (l' +1)(\chi +m-m' )!}} \times \\
   \times \left(\begin{array}{ccc}\hfill l\hfill & \hfill l' \hfill & \hfill \chi \hfill \\ \hfill 0\hfill & \hfill 0\hfill & \hfill 0\hfill \end{array}\right) 
   \left(\begin{array}{ccc}\hfill l\hfill & \hfill l' \hfill & \hfill \chi -1\hfill \\ \hfill m\hfill & \hfill -m' \hfill & \hfill m' -m\hfill \end{array}\right)
   \big[l(l+1)+l' (l' +1)-\chi (\chi +1)\big] {h}_{\chi }^{(1)}(kr_{S,S'}){P}_{\chi }^{|m-m'|}(\cos\theta_{S,S'}),
\end{multline}
otherwise for $\eta\neq\eta'$
\begin{multline}
   W_{d,n,n'}^{S,S'}=e^{i(m-m')\phi_{S,S'}}{(-1)}^{m+m' }\frac{1}{4}\sqrt{\epsilon_{m}\epsilon_{m'}} 
   {\sum }_{\chi =|l-l' |+1}^{l+l' }{\mathrm{i}}^{l' -l+\chi +1}(2\chi +1) 
   \sqrt{\frac{(2l+1)(2l' +1)(\chi -(m-m' ))!}{l(l+1)l' (l' +1)(\chi +m-m' )!}} \times \\ 
   \times \left(\begin{array}{ccc}\hfill l\hfill & \hfill l' \hfill & \hfill \chi -1\hfill \\ \hfill 0\hfill & \hfill 0\hfill & \hfill 0\hfill \end{array}\right)
   \left(\begin{array}{ccc}\hfill l\hfill & \hfill l' \hfill & \hfill \chi \hfill \\ \hfill m\hfill & \hfill -m' \hfill & \hfill m' -m\hfill \end{array}\right)
   \sqrt{\big(\chi^{2}-{(l-l' )}^{2}\big)\big({(l+l' +1)}^{2}-\chi^{2}\big)} 
   {h}_{\chi }^{(1)}(kr_{S,S'}){P}_{\chi }^{|m-m'|}(\cos\theta_{S,S'} ),
\end{multline}
\end{widetext}
where $\epsilon_m\equiv2-\delta_{m0}$ denotes the Neumann symbol.

\section{\corr{Pair correlation function of amorphous arrays\label{ap:pcf}}}
\begin{figure}
    \centering
    \includegraphics{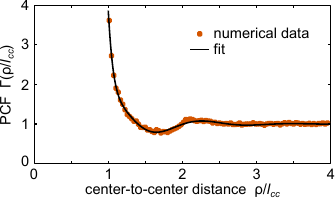}
    \caption{(Color online) Pair correlation function for 2D amorphous arrays are obtained from fitting numerical data.}
    \label{fig:pcf}
\end{figure}

\noindent
\corr{Amorphous arrays are constructed using the RSA algorithm \cite{Hinrichsen1986}, in which the minimum center-to-center distance is a hard limit on particle separation. While the individual neighborhood of each particle is random and hence unique, statistically the density distribution of particles from any one particle is given by a radial PCF \cite{2000_CollSurf_165_287_talbot}. For two dimensional systems the PCF is obtained numerically \cite{Adda-Bedia2008, Bernard2009}, as shown in Fig.~\ref{fig:pcf} with circles, to which we fit an analytical function to allow for easy integration. The numerical PCF data are well fitted by \cite{OE_22_2031_anto}}
\begin{equation}
    \Gamma(x)=1+\sin\left(2\pi\frac{x-x_0}{x_s}\right)\sum_{i=1}^{2}a_{i}e^{-b_{i}(x-x_{i})}
\end{equation}
\corr{with $x=\rho/l_{cc}$ being the CC distance in units of the minimum CC distance. The fitting parameters are $x_0=0.79$, $x_s=1.22$, $a_1=1.02$, $a_2=0.77$, $b_1=17.5$, $b_2=1.62$, $x_1=1.05$, $x_2=0.87$.}

\section{Averaging procedure \label{ap:avg}}
\corr{The passage from Eq. (\ref{eq:before_avg}) to Eq. (\ref{eq:after_avg}) requires averaging over the scatterers in an array and replacing their discrete spatial distribution with continuous one. Here, we summarize the necessary steps and approximations leading to the film of multipoles model.
If we want to find the average scattering coefficients we have to deal with the coupling term from Eq. (\ref{eq:before_avg}) averaged over index $S$, $\frac{1}{N}   \sum_{S} \sum_{S' \neq S} W_{S,S'}(\rho_{S,S'},\phi_{S,S'}) \bm{b^{S'}}$,
where $W_{S,S'}(\rho_{S,S'},\phi_{S,S'})=W_{d}+  W_{r}$.
This term is problematic, because we cannot calculate $b_{S'}$ without solving the equation system or approximations. We assume that the neighbourhood of the particle with index $S$ is composed of approximately identical multipoles, which can be considered a mean field approach. Then, assuming that the array is sufficiently large and uniform, every particle can be approximately described with the same set of multipole moments:}
\begin{multline}
  \frac{1}{N}  \sum_{S} \sum_{S' \neq S} W_{S,S'}(\rho_{S,S'},\phi_{S,S'}) \bm{b^{S'}} = \\
  = \frac{1}{N}  <\bm{b}>  \sum_{S} \sum_{S' \neq S} W_{S,S'}(\rho_{S,S'},\phi_{S,S'}).
\end{multline}
\corr{Because of the fact that $W_{S,S'}$ depends on the position difference, one index can be omitted and the sum can be rewritten in integral form}
\begin{multline}
   \frac{1}{N}  <\bm{b}>  \sum_{S} \sum_{S' \neq S} W_{S,S'}(\rho_{S,S'},\phi_{S,S'}) = \\  <\bm{b}>  \int d^2 r W(\vec{r})\sum_{S''}\frac{\delta(\vec{r}-\vec{r_s''})}{N} .
\end{multline}
\corr{The Dirac $\delta$ term is the local number density of the particles at a certain distance and angle. The local number density can be averaged over disorder realizations as $\sigma \Gamma(\rho/l_{cc},\phi)$, where $\sigma$ is the global particle number density $\Gamma(\rho/l_{cc},\phi)$ is the pair correlation function leading to Eq. \ref{eq:after_avg} with $\bm{b^{S'}}=<\bm{b}>$.}

\section{Relationship between the coupling matrix and the retarded multipole potentials \label{ap:relationship}}
\noindent
Previously, we utilized a Green function based approach to calculate the effective multipole moments (polarizabilities) of particles in an amorphous array, denoted as $S$ \cite{Antosiewicz2012, OE_22_2031_anto, Czajkowski2020, Czajkowski2020a}. One can convert the herein presented result from the VSWF basis back to the multipole moments representation using relations between scattering VSWF expansion coefficients and multipole moments from the literature \cite{Muhlig2011,Grahn2012}.
For magnetic or electric dipoles this relation is expressed as \cite{Muhlig2011}
\begin{equation}
\begin{pmatrix}
  p_x  \\ p_y \\ p_z
 \end{pmatrix} = c 
 \begin{pmatrix}
  b_{1}-b_{-1} \\ i(b_{1}+b_{-1}) \\ -\sqrt{2} b_{0}
 \end{pmatrix}, 
 \end{equation}
where $c$ is a factor required to obtain correct units and wavelength dependence and is defined as $c=-\frac{6i}{4k^3}$ (the difference between \cite{Muhlig2011} and this work stems from CGS units in \cite{Czajkowski2020a}). 
This equation is easily rewritten into matrix form,
\begin{equation}
\begin{pmatrix}
  p_x  \\ p_y \\ p_z
 \end{pmatrix} = c M
 \begin{pmatrix}
  b_{-1,1} \\ b_{0,1} \\ b_{1,1}
 \end{pmatrix},  
 \end{equation}
with $M$ being defined as
  \begin{equation}
  M=
\begin{pmatrix}
  1 & 0 & -1 \\
  i & 0 & i \\
  0 & -\sqrt{2} & 0
 \end{pmatrix}.
 \end{equation}
One can then utilize $M$ for basis conversion
  \begin{equation}
     S=c^{-1} M W M^{-1}.
     \label{eq:swrelation}
 \end{equation}
Finally, we relate $S_{xx}$, which is used in \cite{Czajkowski2020}, to $W$, which is the dipole-dipole part of the coupling matrix used here, by evaluating it with Eq. \ref{eq:swrelation}
\begin{equation}
     S_{xx}=\frac{1}{2c} (W_{-1,-1}+W_{1,1}),
\end{equation}
which, knowing that $W_{-1,-1}=W_{1,1}$, reduces to
\begin{equation}
    S_{xx}=\frac{4ik_0^3}{6} W_{1,1}.
\end{equation}
\acknowledgements
\noindent
We acknowledge support from the Polish National Science Center via the projects 2017/25/B/ST3/00744 and 2019/34/E/ST3/00359. Computational resources were provided by the ICM-UW (Grant~\#G55-6). K.M.C. would like to thank Amos Egel for many fruitful discussions on the T-matrix method and the SMUTHI code.


%

\end{document}